\documentclass[twocolumn]{aastex61}

\usepackage{booktabs}
\received{December 17, 2017}
\revised{xxxx xx, 2018}
\accepted{xxxx xx, 2018}
\submitjournal{ApJ}

\shorttitle{Planet candidates to six northern circumpolar stars}
\shortauthors{G. Jeong et al.}

\begin{document}

\title{A Search for Exoplanets around Northern Circumpolar Stars. IV. Six Planet Candidates to the K Giants, HD 44385, HD 97619, HD 106574, HD 118904, \mbox{HD 164428}, and \mbox{HD 202432}}

\correspondingauthor{Byeong-Cheol Lee}
\email{bclee@kasi.re.kr}

\author{Gwanghui Jeong}
\email{tlotv@kasi.re.kr}
\affiliation{Korea Astronomy and Space Science Institute, 776, Daedeokdae-Ro, Youseong-Gu, Daejeon 34055, Korea}
\affiliation{Korea University of Science and Technology, 217, Gajeong-ro Yuseong-gu, Daejeon 34113, Korea}

\author{Inwoo Han}
\affiliation{Korea Astronomy and Space Science Institute, 776, Daedeokdae-Ro, Youseong-Gu, Daejeon 34055, Korea}
\affiliation{Korea University of Science and Technology, 217, Gajeong-ro Yuseong-gu, Daejeon 34113, Korea}

\author{Myeong-Gu Park}
\affiliation{Department of Astronomy and Atmospheric Sciences, Kyungpook National University, Daegu 41566, Korea}

\author{Artie P. Hatzes}
\affiliation{Th\"uringer Landessternwarte Tautenburg, Sternwarte 5, D-07778 Tautenburg, Germany }

\author{Tae-Yang Bang}
\affiliation{Department of Astronomy and Atmospheric Sciences, Kyungpook National University, Daegu 41566, Korea}

\author{Shenghong Gu}
\affiliation{Yunnan Observatories, Chinese Academy of Sciences, Kunming 650216, China}
\affiliation{University of Chinese Academy of Sciences, Beijing 100049, China}
\affiliation{Key Laboratory for the Structure and Evolution of Celestial Objects, Chinese Academy of Sciences,Kunming 650011, China}

\author{Jinming Bai}
\affiliation{Yunnan Observatories, Chinese Academy of Sciences, Kunming 650216, China}
\affiliation{Key Laboratory for the Structure and Evolution of Celestial Objects, Chinese Academy of Sciences,Kunming 650011, China}

\author{Byeong-Cheol Lee}
\affil{Korea Astronomy and Space Science Institute, 776, Daedeokdae-Ro, Youseong-Gu, Daejeon 34055, Korea}
\affiliation{Korea University of Science and Technology, 217, Gajeong-ro Yuseong-gu, Daejeon 34113, Korea}

\begin{abstract}
We report the discovery of long-period radial velocity (RV) variations in six intermediate mass K giant stars using precise RV measurements. 
These discoveries are part of the Search for Exoplanets around Northern Circumpolar Stars (SENS) survey being conducted at the Bohyunsan Optical Astronomy Observatory (BOAO). 
The nature of the RV variations was investigated by looking for photometric and line shape variations. 
We can find no variability with the RV period in these quantities and conclude that  RV variations are most likely due to unseen sub-stellar companions.
Orbital solutions for the six stars yield orbital periods in the range 418 -- 1065 d and minimum masses in the range 1.9 -- 8.5 $M_{J}$. 
These properties are typical on planets around intermediate mass stars. 
 Our SENS survey so far has about an 8\% confirmed planet occurrence rate, and it will provide better statistics on planets around giant stars when the survey is completed.

\end{abstract}


\section{Introduction} \label{sec:intro}

The measurement of stellar radial velocity (RV) is one of the most effective techniques employed in the search for exoplanets. 
At the Bohyansan Optical Astronomical Observatory (BOAO), we have been conducting an exoplanet search program around
late-type giant stars since 2004. This program has made contributions to both exoplanet and asteroseismic studies around 
K giant stars \citep{2008JKAS...41...59H,2010A&A...509A..24H,2011A&A...529A.134L,2012A&A...546A...5L,2012A&A...548A.118L,2014A&A...566A..67L} and exoplanet detection around G giant stars \citep{2009PASJ...61..825O, 2012PASJ...64...34O, 2013PASJ...65...85S}.

In 2010, we began a new program, the Search for Exoplanet around Northern circumpolar Stars (SENS; \citealt{2015A&A...584A..79L}).
The main goal of SENS is to observe stars that are accessible year-round in order to have better sampling for our targets and thus increase the planet detection efficiency.
The stars of SENS were selected
 from the \textit{HIPPARCOS} catalog with visual magnitudes of 5.0 $<$ $m_{v}$ $<$ 7.0 and 
color indices of 0.6 $<$ $\textit{B -- V}$ $<$ 1.6.
The original SENS sample consist of 224 stars -- 5\% dwarfs, 40\% giant stars, and 55\% unclassified stars.
From SENS survey, we detected periodic RV variations around roughly twenty G, K, and M giant stars.
Among them, \mbox{HD 104985} \citep{2003ApJ...597L.157S}, 11 Ursae Minoris \citep{2009A&A...505.1311D}, \mbox{HD208527} \& \mbox{HD 220074} \citep{2013A&A...549A...2L}, \mbox{HD 11755}, \mbox{HD 12648}, \mbox{HD 24064} and 8 Ursae Minoris \citep{2015A&A...584A..79L}, \mbox{HD 36384}, \mbox{HD 52030}, and \mbox{HD 208742} \citep{2017ApJ...844...36L} were later
shown to host planetary companions.

In this paper, the observational strategy and data reduction are summarized in Section 2. 
Section 3 describes the stellar properties and analysis of each host stars. 
In Section 4, orbital solutions are described in detail. 
Some possible causes of the RV variations are examined in Section 5.
The discussion about the results is presented in Section 6.

\section{Observation} \label{sec:obs}

Observations were made with the high-resolution fiber-fed Bohyunsan Observatory Echelle Spectrograph (BOES; \citealt{2007PASP..119.1052K}) of the 1.8 m telescope at BOAO. 
An iodine (I$_{2}$) absorption cell is equipped in BOES for precise RV measurements.
BOES has three fibers with 80 $\micron$, 200 $\micron$, and 300 $\micron$ of diameter, which give resolutions of $R$ = 90,000, 45,000, and 27,000, respectively.

For our program, we used the 200 $\micron$ fiber. A  typical exposure time of 20 minutes
 yielded a signal-to-noise ratio (S/N) of about 150.
Since the start of the program in January 2010, we have collected 30 spectra each
for \mbox{HD 44385}, \mbox{HD 97619}, \mbox{HD 106574}, \mbox{HD 118904}, \mbox{HD 164428}, and \mbox{HD 202432}.
In order to check the instrumental stability, we have monitored an $\tau$ Ceti by standard since 2003.
The long-term RV accuracy of BOES is $\sim$7.6 m s$^{-1}$.

The data reduction was performed with the IRAF package for bias subtraction, flat fielding, and oder extraction, etc. 
Once we extracted the 1-D spectra from the raw data,  precise RVs were measured
using the program RVI2CELL \citep{2007PKAS...22...75H}.
Tables~\ref{tab:rv1}--\ref{tab:rv6} list the measured RVs of each star.

\section{Stellar characteristics} \label{sec:ste}

We obtained the basic stellar parameters (the visual magnitude V, parallax $\pi$, spectral type, B-V, luminosity, and RV) of the stars from \citet{2012AstL...38..331A} based on the \textit{HIPPARCOS} catalog \citep{1997yCat.1239....0E}. We also adopted more precise parallaxes from \citet{2016A&A...595A...1G}.
The effective temperature, log \textit{g}, metalicity ([Fe/H]), and microturbulent velocity ($v_{\rm{micro}}$) were estimated from TGVIT \citep{2005PASJ...57...27T}.
For comparison, we also obtained the effective temperature from \citet{2006ApJ...638.1004A} and \citet{2012MNRAS.427..343M,2017MNRAS.471..770M}.
We have estimated the projected rotational velocities using a line broadening technique \citep{2008PASJ...60..781T}. 

The stellar radii, masses, and ages were calculated using the online tool PARAM 1.3\footnote{\url{http://stev.oapd.inaf.it/cgi-bin/param_1.3/}} \citep{2006A&A...458..609D}, which is based on a library of theoretical stellar isochrones \citep{2000A&AS..141..371G,2012MNRAS.427..127B} and Bayesian inference \citep{2005A&A...436..127J}.

\begin{table*}[h!]
\renewcommand{\thetable}{\arabic{table}}
\centering
\caption{Stellar parameters for the stars.} \label{tab:ste}
\begin{tabular}{cccccccc}
\toprule[1.5pt]
	Parameter	& HD 44385	& HD 97619	& HD 106574	& HD 118904	& HD 164428	& HD 202432	& Ref.\\
\midrule
Spectral type			& K0	& K0	& K2 III	& K2 III	& K5	& K2	& 1\\
$\textit{$m_{v}$}$ (mag)	& 6.771 $\pm$ 0.001	& 7.044 $\pm$ 0.001	& 5.883 $\pm$ 0.001	& 5.665 $\pm$ 0.001	& 6.388 $\pm$ 0.001	& 7.052 $\pm$ 0.001	& 1\\
$\textit{B -- V}$ (mag)	& 1.266 $\pm$ 0.007 	& 1.315 $\pm$ 0.008	& 1.179 $\pm$ 0.005	& 1.219 $\pm$ 0.005 	& 1.452 $\pm$ 0.008	& 1.206 $\pm$ 0.009	& 1\\
RV (km s$^{-1}$)		& $-$ 14.87 $\pm$ 0.20	& $-$ 23.84 $\pm$ 0.17	& $-$ 16.48 $\pm$ 0.29	& 13.72 $\pm$ 0.20	& $-$ 8.16 $\pm$ 0.20	& $-$ 2.23 $\pm$ 0.20	&  1\\
$\pi$ (mas)			& 3.92 $\pm$ 0.41	& 4.93 $\pm$ 0.42	& 7.00 $\pm$ 0.28	& 7.93 $\pm$ 0.24	& 3.82 $\pm$ 0.29	& 6.40 $\pm$ 0.40	& 1\\
					& 4.68 $\pm$ 0.29	& 4.69 $\pm$ 0.33	& --	& 9.02 $\pm$ 0.76	& 3.65 $\pm$ 0.27	& 6.20 $\pm$ 0.23	& 4\\
$T_{\rm{eff}}$ (K)		& 4499	& 4245	& --	& 4511	& 4222	& 4569	& 2\\
				& 4326	& 4334	& 4482	& 4424	& 4257	& 4465	& 3\\
				& 4433 $\pm$ 125	& 4237 $\pm$ 125	& 4414 $\pm$ 125	& 4407 $\pm$ 125	& 4109 $\pm$ 125	& 4458 $\pm$ 125	& 5\\
				& 4440 $\pm$ 28	& 4355 $\pm$ 20	& 4501 $\pm$ 33	& 4469 $\pm$ 23	& 4119 $\pm$ 40	& 4549 $\pm$ 30	& 6\\
$\rm{[Fe/H]}$		& 0.10 $\pm$ 0.07	& $-$ 0.07 $\pm$ 0.07	& $-$ 0.43 $\pm$ 0.04	& $-$ 0.11 $\pm$ 0.09	& $-$ 0.07 $\pm$ 0.10	& 0.16 $\pm$ 0.10	& 6\\
log $\it g$ (cgs)		& 1.80	& 1.83	& 1.78	& 1.91	& 1.21	& 2.24	& 5\\
				& 2.00 $\pm$ 0.12	& 2.33 $\pm$ 0.09	& 2.18 $\pm$ 0.18	& 2.13 $\pm$ 0.10	& 1.62 $\pm$ 0.17	& 2.42 $\pm$ 0.12	& 6\\
$v_{\rm{micro}}$ (km s$^{-1}$)	& 1.56 $\pm$ 0.13	& 1.54 $\pm$ 0.13	& 1.59 $\pm$ 0.06	& 1.39 $\pm$ 0.15	& 1.62 $\pm$ 0.16	& 1.32 $\pm$ 0.16	& 6\\
Age (Gyr)		& 1.8 $\pm$ 0.4	& 4.9 $\pm$ 1.7	& 4.6 $\pm$ 1.3	& 3.7 $\pm$ 1.6	& 2.7 $\pm$ 1.2	& 6.1 $\pm$ 2.6	& 6\\
$\textit{$R_{\star}$}$ ($R_{\odot}$)	& 19.5	& 19.5	& 20.4	& 17.7	& 39.4	& 12.3	& 5\\
							& 18.2 $\pm$ 1.2	& 16.7 $\pm$ 1.3	& 17.1 $\pm$ 0.9	& 14.8 $\pm$ 1.3	& 35.4 $\pm$ 2.9	& 11.1 $\pm$ 0.3	& 6\\
$\textit{$M_{\star}$}$ ($M_{\odot}$)	& 1.8 $\pm$ 0.2	& 1.3 $\pm$ 0.1	& 1.2 $\pm$ 0.1	& 1.4 $\pm$ 0.2	& 1.5 $\pm$ 0.2	& 1.2 $\pm$ 0.2	& 6\\
$\textit{$L_{\star}$}$ ($L_{\odot}$)	& 233.2	& 122.8	& 152.2	& 152.4	& 469.3	& 63.9		& 1 \\
							& 132.6 $\pm$ 11.1	& 109.8 $\pm$ 10.1	& 142.5 $\pm$ 9.8	& 105.6 $\pm$ 10.7	& 397.9 $\pm$ 37.8	& 54.0 $\pm$ 3.6	& 5\\
							& 116.0 $\pm$ 15.6	& 90.4 $\pm$ 14.2	& 108.1 $\pm$ 11.8	& 78.7 $\pm$ 13.9	& 325.0 $\pm$ 54.7	& 47.5 $\pm$ 2.9	& 6\\
$v_{\rm{rot}}$ sin $i$ (km s$^{-1}$)	& 2.6 $\pm$ 0.5	& 1.9 $\pm$ 0.5	& 1.7 $\pm$ 0.5	& 1.2 $\pm$ 0.5	& 2.8 $\pm$ 0.5	& 2.1 $\pm$ 0.5	& 6\\
$P_{\rm{rot}}$ / sin $i$ (days)		& 354.1 $\pm$ 72.0	& 444.7 $\pm$ 122.0	& 508.9 $\pm$ 152.1	& 624.0 $\pm$ 265.7	& 629.6 $\pm$ 125.7	& 267.4 $\pm$ 64.1	& 6\\
\bottomrule[1.5pt]
\end{tabular}
\textbf{References.}--- (1)  Anderson \& Francis (2012); (2) McDonald et al (2012); (3) Ammons et al (2006); \\ (4) Gaia Collaboration et al (2016); (5) McDonald et al (2017); (6) This work. \\
\end{table*}

The stellar parameters of the observed stars are summarized in Table~\ref{tab:ste}.

\section{Orbital solutions} \label{sec:orb}

\begin{table*}[h!]
\renewcommand{\thetable}{\arabic{table}}
\centering
\caption{Preliminary orbital solutions.} \label{tab:orb}
\begin{tabular}{lcccccc}
\toprule[1.5pt]
    Parameter	& HD 44385 b & HD 97619 b & HD 106574 b & HD 118904 b & HD 164428 b & HD 202432 b\\
\midrule
P (days)	& 473.5 $\pm$ 4.9	& 665.9 $\pm$ 9.5	& 1065.7 $\pm$ 14.6	& 676.7 $\pm$ 19.1	& 599.6 $\pm$ 8.7	&  418.8 $\pm$ 2.9 \\
K (m s$^{-1}$)	& 104 $\pm$ 10	& 68 $\pm$ 6	& 149 $\pm$ 8	& 61 $\pm$ 8	& 109 $\pm$ 15	& 43 $\pm$ 3   \\
$T_{periastron}$ (JD)  & 2455121 $\pm$ 37	& 2455238 $\pm$ 55	& 2455585 $\pm$ 360	& 2455478 $\pm$ 82	& 2455068 $\pm$ 46	& 2454908 $\pm$ 26\\ 
$e$			& 0.20 $\pm$ 0.20	& 0.23 $\pm$ 0.17	& 0.03 $\pm$ 0.03	& 0.31 $\pm$ 0.30	& 0.29 $\pm$ 0.22	& 0.21 $\pm$ 0.16  \\
$\omega$ (deg)  & 292.18 $\pm$ 28.11 & 293.53 $\pm$ 23.97	& 44.66 $\pm$ 122.15	& 29.16 $\pm$ 31.02	& 203.11 $\pm$ 26.59	& 61.24 $\pm$ 21.76  \\
$m$ sin $i$ ($M_{J}$) & 5.9 $\pm$ 1.1	& 3.5 $\pm$ 1.3	& 8.5 $\pm$ 1.1	& 3.1 $\pm$ 1.2	& 5.7 $\pm$ 1.3	& 1.9 $\pm$ 0.4\\
$a$ (AU)           & 1.4 $\pm$ 0.1	& 1.6 $\pm$ 0.1	& 2.2 $\pm$ 0.1	& 1.7 $\pm$ 0.1	& 1.6 $\pm$ 0.1	& 1.2 $\pm$ 0.1\\
Slope (m s$^{-1} \rm{yr}^{-1}$)	& -6.9	& 11.5	& 2.4	& 1.0	& 1.2	& 4.4	\\
$N_{obs}$        & 35	& 35	& 31	& 38	& 30	& 29\\
rms (m s$^{-1}$)	& 41.5	& 26.2	& 44.0	& 33.2	& 51.6	& 12.4\\
\bottomrule[1.5pt]
\end{tabular}
\end{table*} 

Here we present the derived orbital parameters assuming that periodic RV variations are caused by Keplerian motion. 
An initial value of the period was determined using the Lomb-Scargle periodogram
(L-S) analysis, which is appropriate for uneven sampled data and also gives an estimate of the false alarm probability (FAP) of the signal \citep{1976Ap&SS..39..447L,1982ApJ...263..835S}. 
Using this initial period, we estimated all the orbital elements by an iterated non-linear least-squares method. 
If a linear RV trend is shown, a slope is taken as an unknown parameter.
Table~\ref{tab:orb} lists the orbital parameters for all stars.

\subsection{HD 44385} \label{subsec:44385}

\begin{figure} [h!]
\plotone{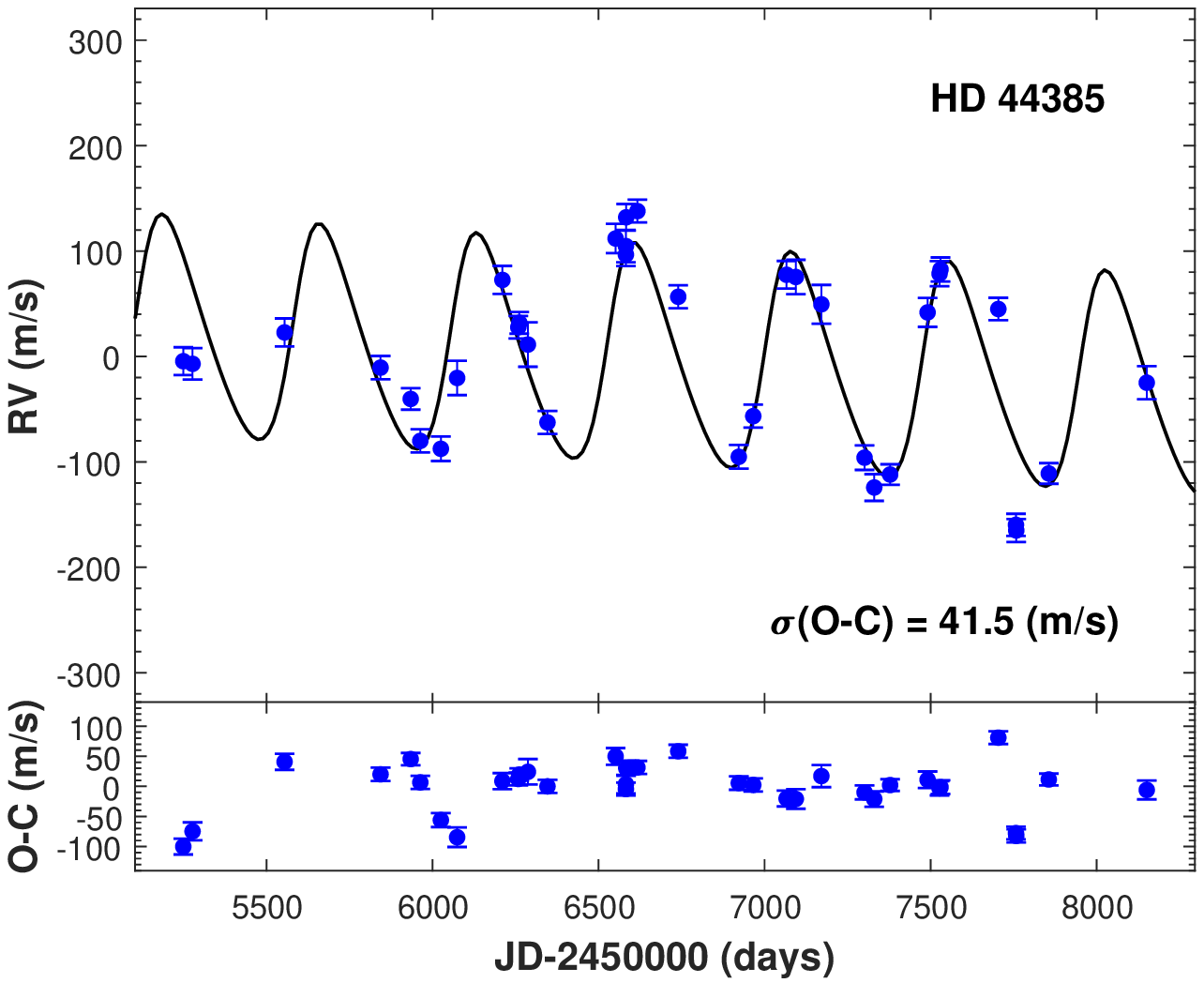} 
\caption{\textit{Upper panel}: RVs of HD 44385 (blue dots) and a fitted Keplerian orbit (solid line). \textit{Lower panel}: The residual RVs after subtracting the Keplerian fit.}
\label{fig:rv44385}
\end{figure}

The RV data of HD 44385 and the Keplerian motion are plotted in Fig.~\ref{fig:rv44385}.
The RVs cover more than five cycles and show a slight linear trend of --6.9 m s$^{-1} \rm{yr}^{-1}$, such linear trend is also seen in the other five systems of this study.
These linear RV variations may be caused by an unseen distant companion or due to some long-term and unknown intrinsic stellar variations.
The orbit has a period, $P$ =  \mbox{473.5 $\pm$ 4.9 days}, an eccentricity, {$e$ = 0.20 $\pm$ 0.20}, and a semi-amplitude $K$ = 104 $\pm$ 10 m s$^{-1}$. 
.

The RV residuals have an rms of 41.5 m s$^{-1}$, which is larger than the typical intrinsic RV variations of K giant stars.
The scaling relationship of \citet{1995A&A...293...87K} yields an amplitued of 15 m s$^{-1}$ for stellar oscillations, which may contribute to rms of the RV residuals.
However, the excess residual RV scatter is still too large.
We discuss these large residual RV scatter in Section \ref{sec:dis}.

\subsection{HD 97619} \label{subsec:97619}

\begin{figure} [h!]
\plotone{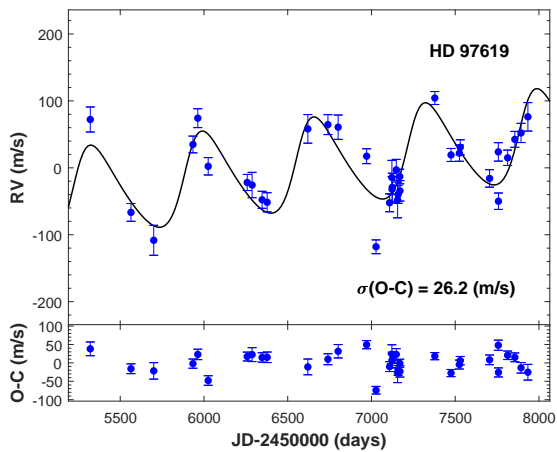}
\caption{\textit{Upper panel}: RVs of HD 97619 (blue dots) and a fitted Keplerian orbit (solid line). \textit{Lower panel}: The residual RVs after subtracting the Keplerian fit.}
\label{fig:rv97619}
\end{figure}

The RV data of HD 97619 and the Keplerian orbit are plotted in Fig.~\ref{fig:rv97619}.
The RVs cover more than three cycles and these also show a slight linear trend of 11.5 m s$^{-1} \rm{yr}^{-1}$, again possibly due  a third body in the system.
The best Keplerian fit yields the orbital elements of $P$ = \mbox{665.9 $\pm$ 9.5 days}, {$e$ = 0.23 $\pm$ 0.17}, and $K$ = 68 $\pm$ 6 m s$^{-1}$.
The host star has a mass just a little higher than the Sun, as the other host star HD 202432. 
It hosts a planet candidate with the lowest mass among the companions in the present study.
The rms of RV residuals is 26.2 m s$^{-1}$ consistent  with the expected RV scatter (so called ``jitter'').

\subsection{HD 106574} \label{subsec:106574}

\begin{figure} [h!]
\plotone{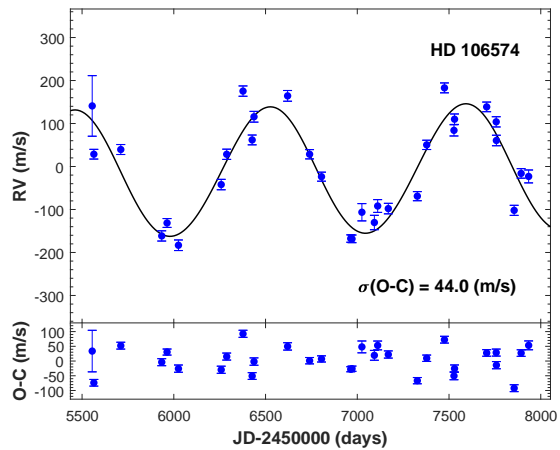}
\caption{\textit{Upper panel}: RVs of HD 106574 (blue dots) and a fitted Keplerian orbit (solid line). \textit{Lower panel}: The residual RVs after subtracting the Keplerian fit.}
\label{fig:rv106574}
\end{figure}

The RV data of HD 106574 and the best Keplerian orbital fit are plotted in Fig.~\ref{fig:rv106574}.
The RVs cover about two cycles.
The orbit has a period of \mbox{1065.7 $\pm$ 14.6 days}, an eccentricity of 0.03 $\pm$ 0.03, and a semi-amplitude of 149 $\pm$ 8 m s$^{-1}$.
Unlike other five systems, this system has a nearly circular orbit.
The planet candidate is the most massive among our six stars.
The residuals have an rms of 44.0 m s$^{-1}$. This is considerably larger than the value of 21 m s$^{-1}$ predicted by the \citet{1995A&A...293...87K} relationship.
This possibly indicates an additional source of stellar variability as HD 44385.

\subsection{HD 118904} \label{subsec:118904}

\begin{figure} [h!]
\plotone{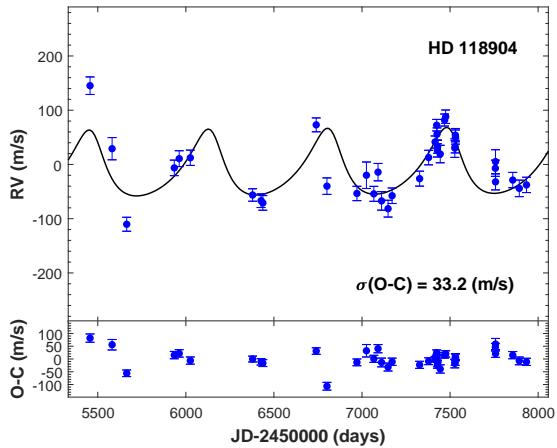}
\caption{\textit{Upper panel}: RVs of HD 118904 (blue dots) and a fitted Keplerian orbit (solid line). \textit{Lower panel}: The residual RVs after subtracting the Keplerian fit.}
\label{fig:rv118904}
\end{figure}

The RV data of HD 118904 and the best Keplerian orbital fit are plotted in Fig.~\ref{fig:rv118904}.
The RVs cover more than three cycles.
The orbital fit yields a period of \mbox{676.7 $\pm$ 19.1 days}, an eccentricity of 0.31 $\pm$ 0.30, and a semi-amplitude of 61 $\pm$ 8 m s$^{-1}$.
The rms RV residuals have a value of 33.2 m s$^{-1}$, also the normal case for a K giant star.

\subsection{HD 164428} \label{subsec:164428}

\begin{figure} [h!]
\plotone{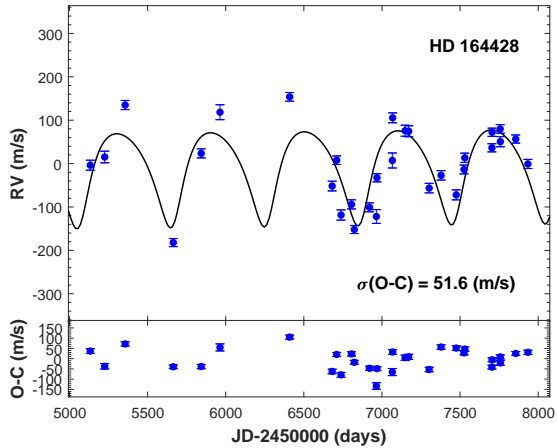}
\caption{\textit{Upper panel}: RVs of HD 164428 (blue dots) and a fitted Keplerian orbit (solid line). \textit{Lower panel}: The residual RVs after subtracting the Keplerian fit.}
\label{fig:rv164428}
\end{figure}

The RV data of HD 164428 and the  Keplerian orbital fit are plotted in Fig.~\ref{fig:rv164428}.
The RVs cover more than four cycles.
The  orbit has a period of \mbox{599.6 $\pm$ 8.7 days},  an eccentricity of 0.29 $\pm$ 0.22, and a semi-amplitude of 109 $\pm$ 15 m s$^{-1}$.
The planet candidate has a similar mass as HD 44385 b.
The rms scatter about the orbit  is 51.6 m s$^{-1}$, which is the largest jitter seen in our six systems, and is consistent with the predicted value of $\sim$50 m s$^{-1}$ for the stellar oscillations according to scaling relationships.

\subsection{HD 202432} \label{subsec:202432}

\begin{figure} [h!]
\plotone{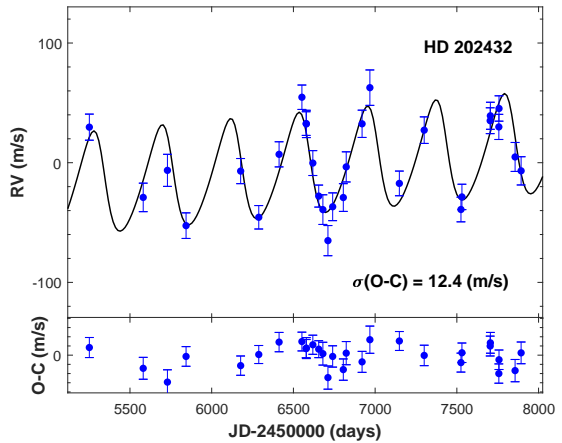}
\caption{\textit{Upper panel}: RVs of HD 202432 (blue dots) and a fitted Keplerian orbit (solid line). \textit{Lower panel}: The residual RVs after subtracting the Keplerian fit.}
\label{fig:rv202432}
\end{figure}

The RV data of HD 202432 and the best Keplerian orbital fit are plotted in Fig.~\ref{fig:rv202432}.
The RVs cover six cycles and show a slight linear trend of 4.4 m s$^{-1} \rm{yr}^{-1}$. 
The Keplerian fit yields a period of \mbox{418.8 $\pm$ 2.9 days}, an eccentricity of 0.21 $\pm$ 0.16, and a semi-amplitude of 43 $\pm$ 3 m s$^{-1}$.
An rms of RV residuals is 12.4 m s$^{-1}$, a value consistent with the expected velocity amplitude of 10 m s$^{-1}$ for stellar oscillations.
\\

\begin{figure*} [h]
\plotone{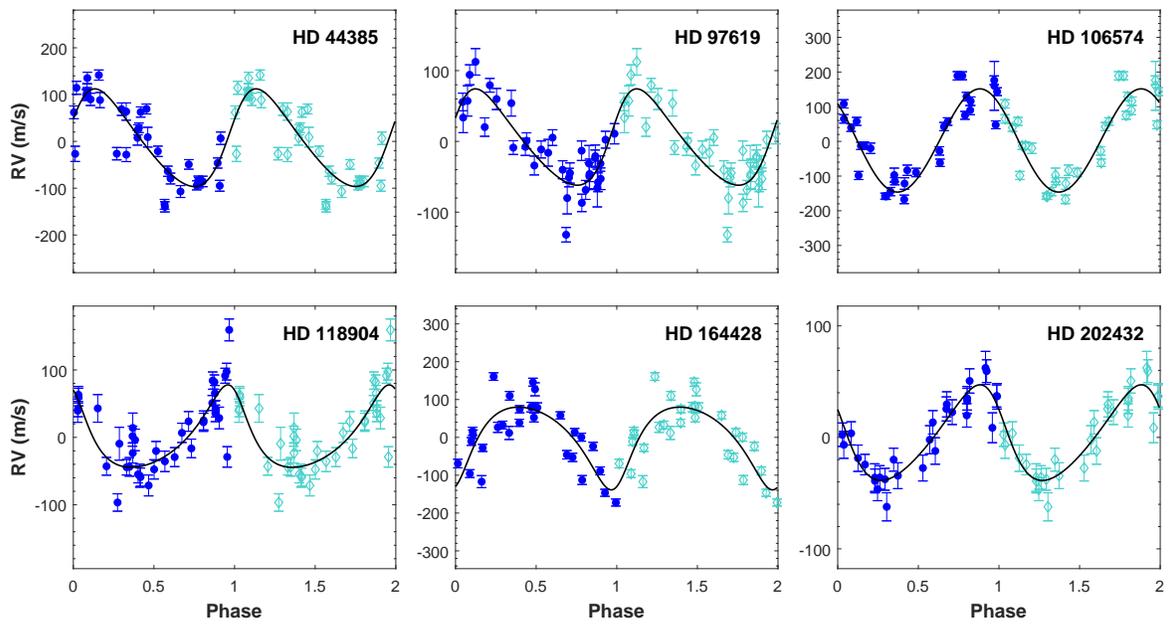}
\caption{The phase folded RV curves of all stars}
\label{fig:pha}
\end{figure*}

The RV measurements phased using best-fit orbital periods for all six stars is shown in Figure~\ref{fig:pha}.
All planet candidates except HD 106574 b have a relatively high eccentricity.
However, we have a relatively small number of observations and these stars show significant intrinsic variability. 
The coarse sampling plus large scatter may result in an artificially high eccentricity. More data are needed to resolve this.

\section{The cause of the RV variations} \label{sec:cau}

The periodic variations of RVs may be also intrinsic to the star such as rotational modulation or stellar pulsations by surface features. 
To identify the nature of the RV variations, we investigated Ca II H lines, photometric data, and spectral line profile variations.

\subsection{Surface activity} \label{subsec:sur}

\begin{figure} [h!]
\plotone{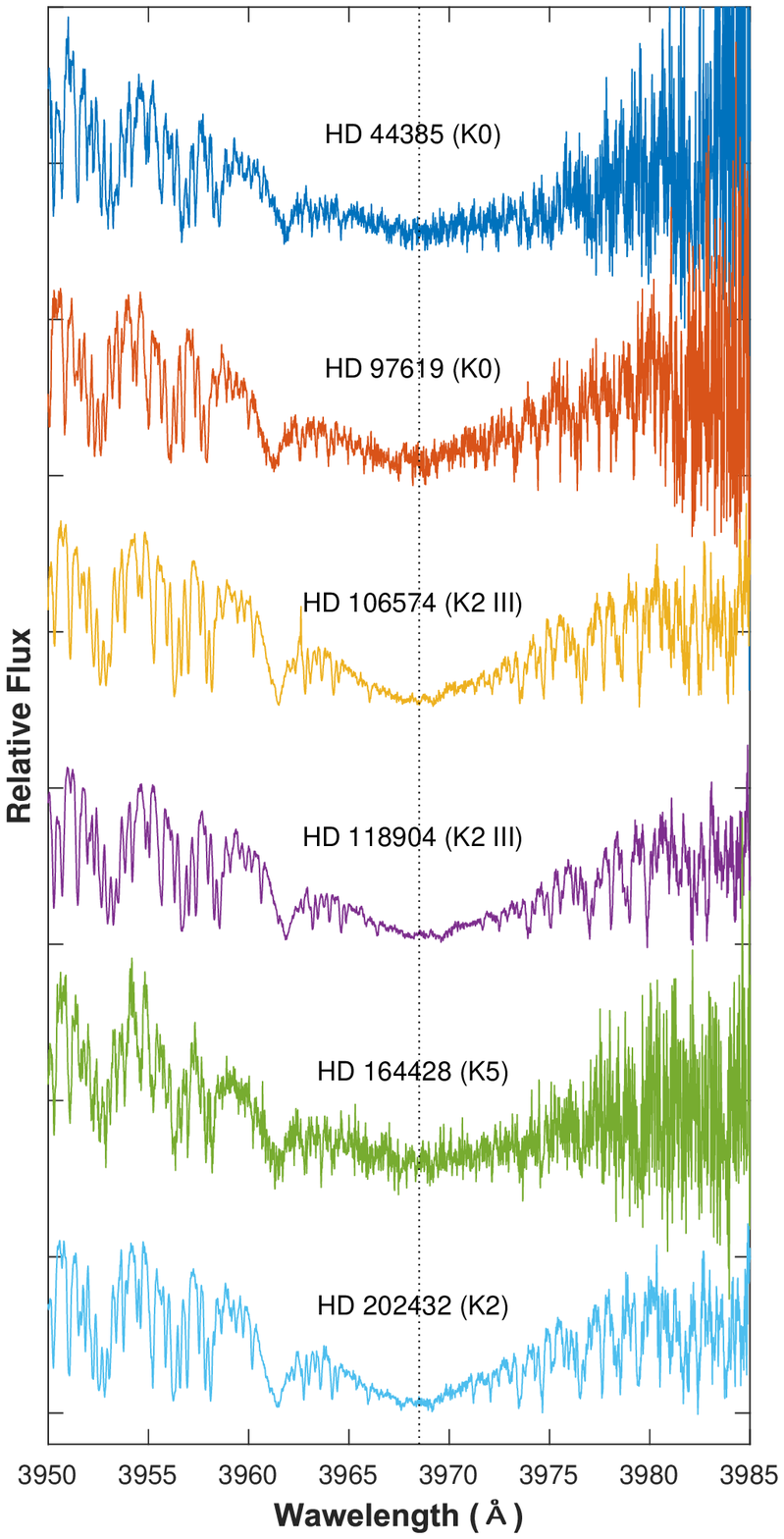}
\caption{Spectra in the region of the Ca II H line. The vertical dotted-line is located in the core of the Ca II H profiles (\mbox{3968.5 $\AA$}).}
\label{fig:ca}
\end{figure}

A rotating star with surface features such as spots or  plage caused by magnetic activity will exhibit periodic RV variations, which can be misinterpreted as a planetary signal \citep{2001A&A...379..279Q}.
The Ca II H line (3968.5 {\AA}) has often been used as an good indicator of stellar activity \citep{1913ApJ....38..292E}. 
If there is a chromospheric activity, an emission line may appear at the center of Ca II H absorption line profile on the core of the line may be partially filled in.

The Ca II H lines of each host star are shown in Fig.~\ref{fig:ca}. 
There appears to be no noteworthy emission in the core of the Ca II H line for any target stars.

\subsection{Photometric variations} \label{subsec:pho}

\begin{figure} [h]
\plotone{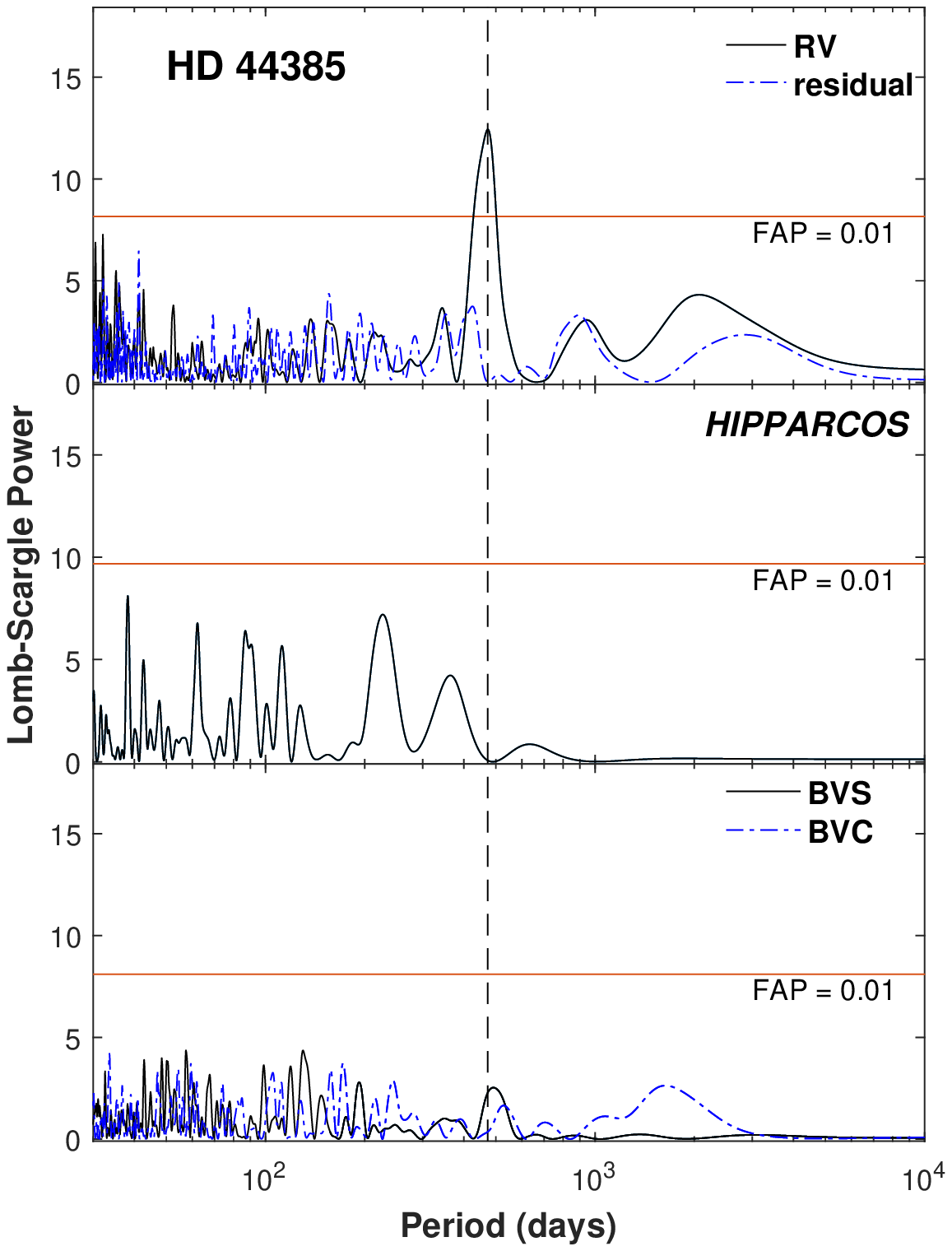}
\caption{\textit{Top to bottom}: L-S periodogram of the RV measurements, residual RVs, \textit{HIPPARCOS} photometry, and the line bisectors of the span and curvature for HD 44385. The vertical dashed  line represents 473 d period.}
\label{fig:ls44385}
\end{figure}

\begin{figure} [h]
\plotone{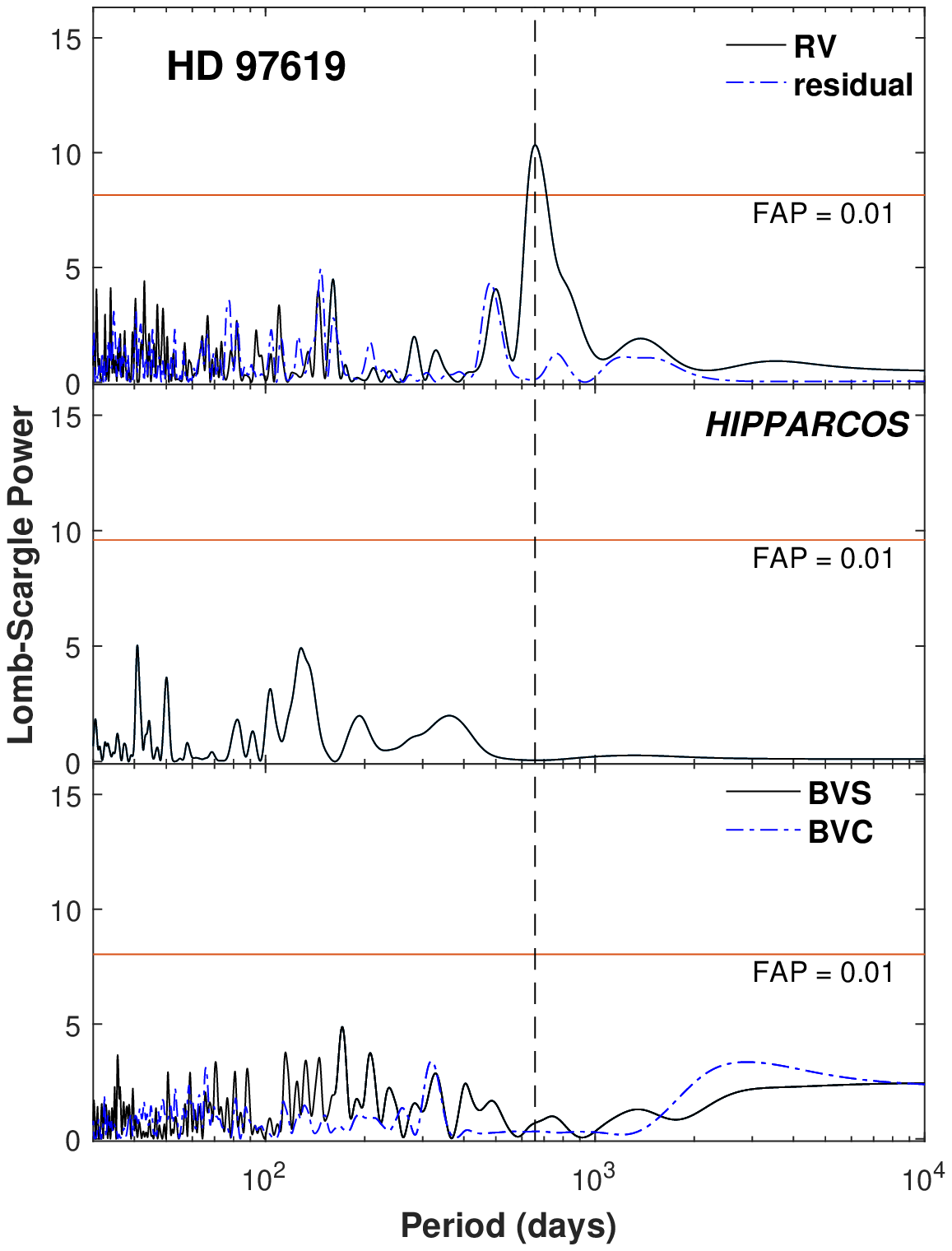}
\caption{\textit{Top to bottom}: L-S periodogram of the RV measurements, residual RVs, \textit{HIPPARCOS} photometry, and the line bisectors of the span and curvature for HD 97619. The vertical dashed line represents 667 d period.}
\label{fig:ls97619}
\end{figure}

\begin{figure} [h]
\plotone{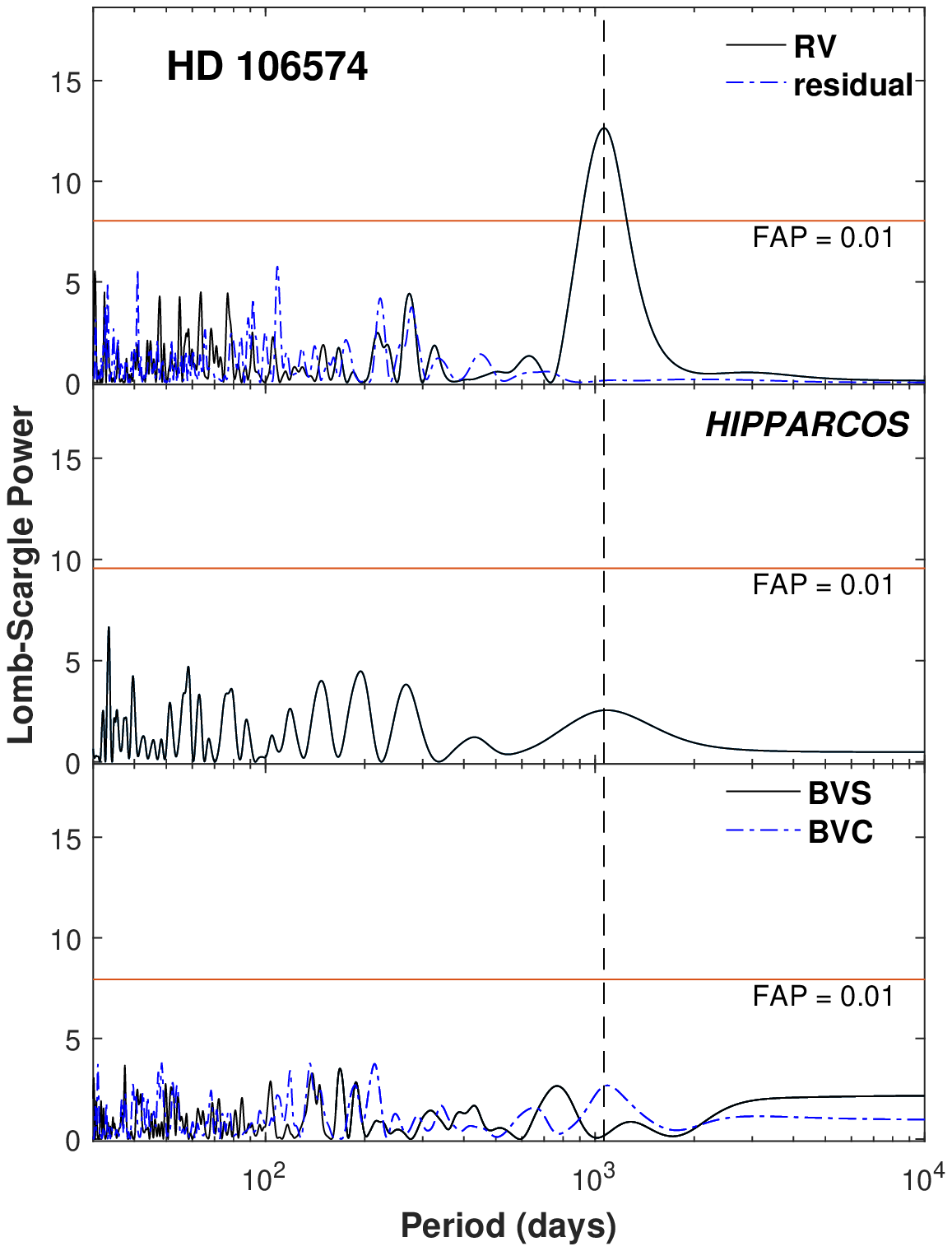}
\caption{\textit{Top to bottom}: L-S periodogram of the RV measurements, residual RVs, \textit{HIPPARCOS} photometry, and the line bisectors of the span and curvature for HD 106574. The vertical dashed line represents 1071 d period.}
\label{fig:ls106574}
\end{figure}

\begin{figure} [h]
\plotone{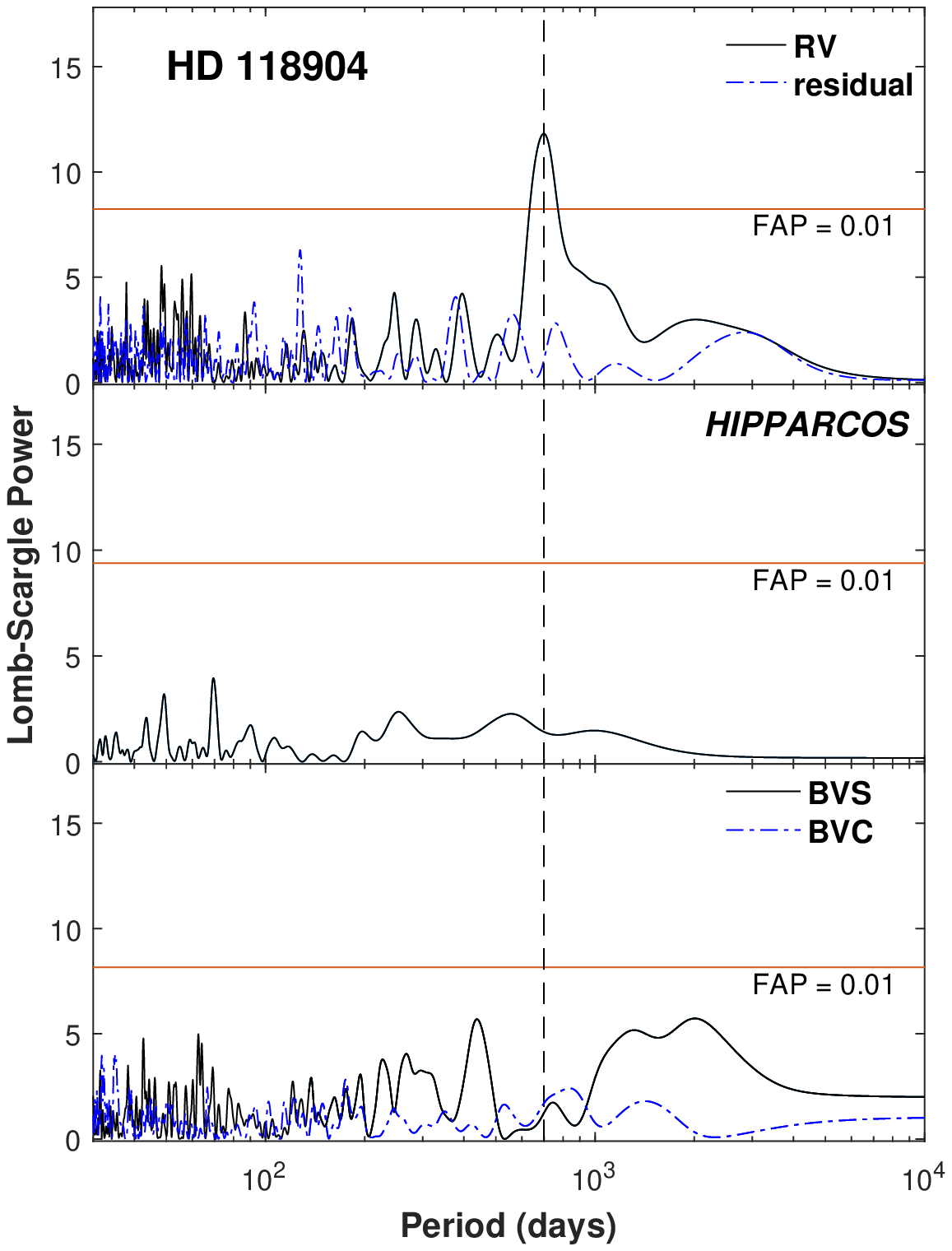}
\caption{\textit{Top to bottom}: L-S periodogram of the RV measurements, residual RVs, \textit{HIPPARCOS} photometry, and the line bisectors of the span and curvature for HD 118904. The vertical dashed line represents 675 d period.}
\label{fig:ls118904}
\end{figure}

\begin{figure} [h]
\plotone{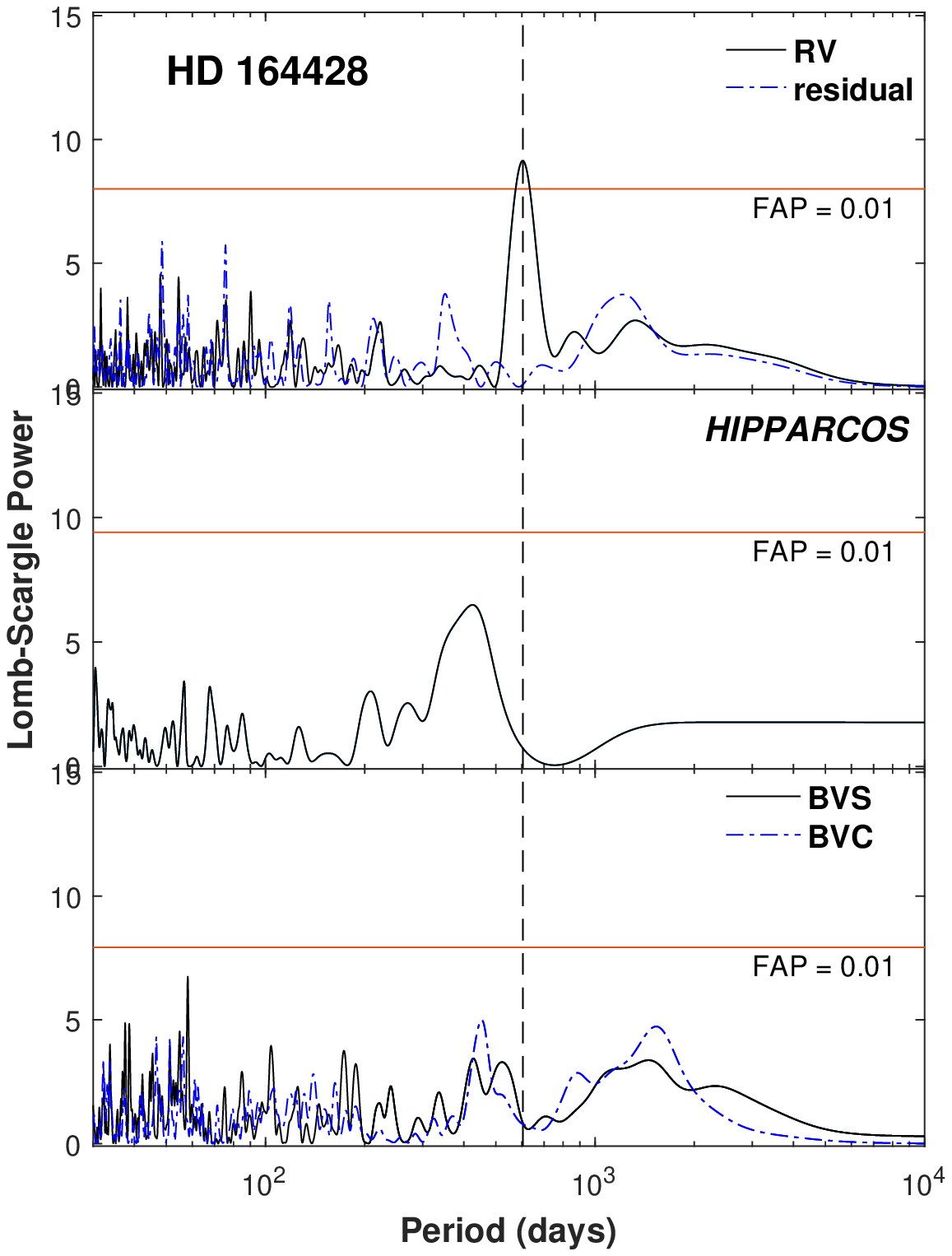}
\caption{\textit{Top to bottom}: L-S periodogram of the RV measurements, residual RVs, \textit{HIPPARCOS} photometry, and the line bisectors of the span and curvature for HD 164428. The vertical dashed  line represents 594 d period.}
\label{fig:ls164428}
\end{figure}

\begin{figure} [h]
\plotone{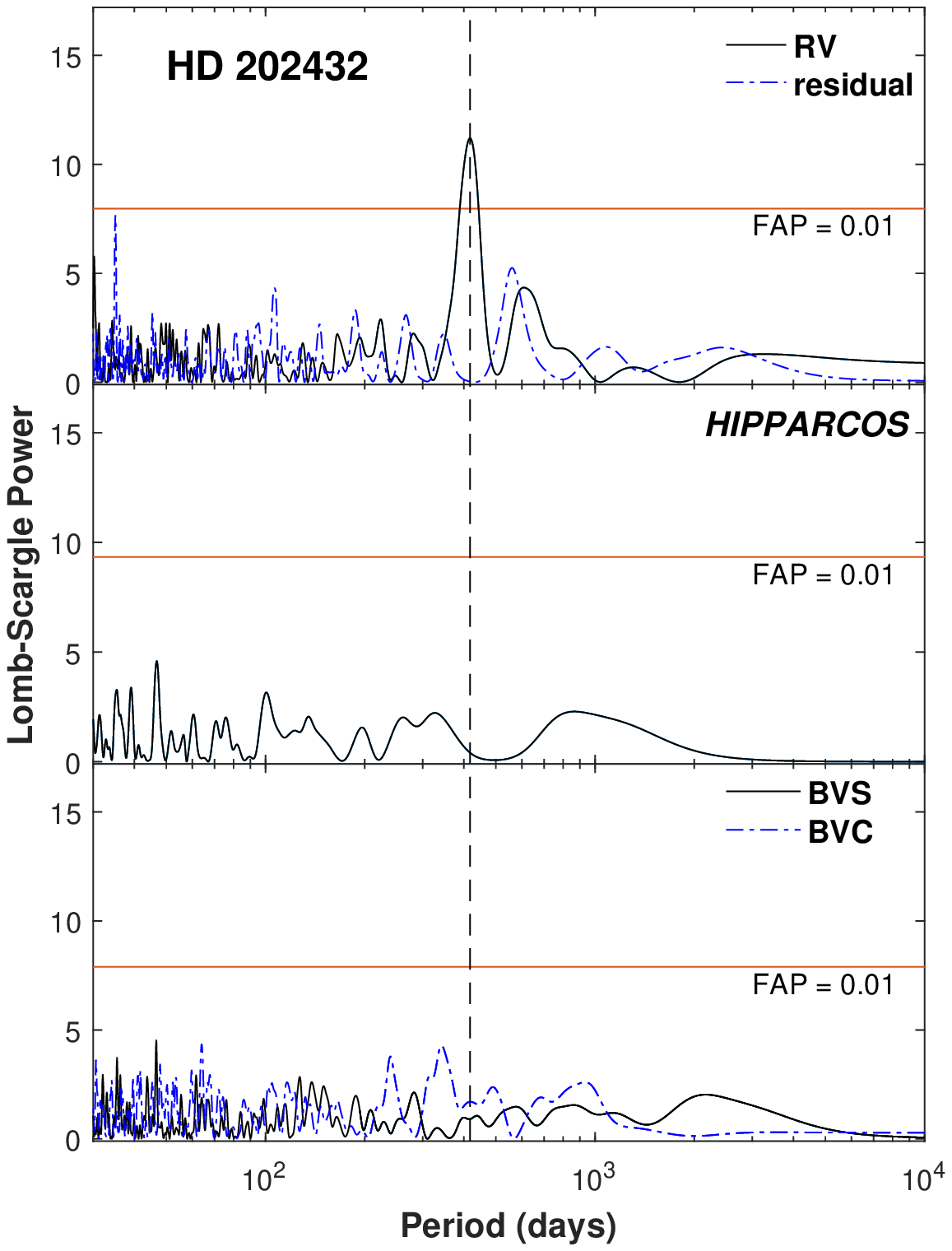}
\caption{\textit{Top to bottom}: L-S periodogram of the RV measurements, residual RVs, \textit{HIPPARCOS} photometry, and the line bisectors of the span and curvature for HD 202432. The vertical dashed  line represents 422 d period.}
\label{fig:ls202432}
\end{figure}



To check for any photometric variations of the stars, we used \textit{HIPPARCOS} archival data.
\textit{HIPPARCOS} data cover the period between November 1989 and March 1993.
Although the measurements are not contemporaneous with our observations, it is still useful to check whether the \textit{HIPPARCOS} data show any periodic photometric variability.
The \textit{HIPPARCOS} catalogue provides a total of 160, 148, 143, 119, 122, and 114 observations
of the \mbox{HD 44385}, \mbox{HD 97619}, \mbox{HD 106574}, \mbox{HD 118904}, \mbox{HD 164428}, and \mbox{HD 202432}, respectively.

The rms scatter of the data is 0.011, 0.011, 0.007, 0.011, 0.009, and 0.010 mag for each star, respectively.
Their scatter is comparable to the photometric uncertainties of \textit{HIPPARCOS} data
which ranges from 0.007 to 0.011 mag.
We do not see any significant photometric variability of the stars.
We also checked periodicity of \textit{HIPPARCOS} data from the L-S periodogram, and could not find any significant periodicity corresponding to that of the RV variations (middle panels in the Figs.~\ref{fig:ls44385}--\ref{fig:ls202432}).

\subsection{Line bisector variations} \label{subsec:lin}

The examination of spectral line shape is an important tool to help identify other origins of RV variations than Keplerian motion \citep{1998ASPC..154..311H}. 
Orbital motion by a companion should cause a Doppler shift without a change of the line shape, whereas RV variations caused by rotational modulation from stellar surface structure should be accompanied by line shape variations. 
We thus investigated the change of the spectral line shapes using two indicators:
The bisector velocity curvature (BVC) and the bisector velocity span (BVS).

To measure the line bisectors, we selected the lines \mbox{Ca I} 6122.2, 6439.1, 6462.6, 6717.7; \mbox{Cr I} 6330.1; \mbox{Fe I} 6141.7, 6393.6, 6411.7, 6677.9, 6750.2; \mbox{Fe II} 6151.6; \mbox{Ni I} 6643.6, 6767.8; \mbox{Ti I} 6085.2, 6742.6; \mbox{V I} 6039.7, 6081.4.
These lines are free from I$_{2}$ and telluric absorption lines and are relatively deep.
The L-S periodogram of bisectors do not show any significant periodicity associated with those of RV variations (bottom panels in the Figs~\ref{fig:ls44385}--\ref{fig:ls202432}).

\section{Discussion} \label{sec:dis}

We have found long-period RV variations of six K-giant stars. 
The stars show no variations in the line shapes as measured by the spectral line bisectors. 
HD~44385 does show a weak signal in the bisector curvature measurements (BVS), but this does not seem to be significant (FAP $\ge$ 0.1). 
We note that a lack of bisector measurements is not proof of the planetary nature. 
High quality bisector measurements are difficult to make as these require high spectral resolution and high S/N data. 
In general these are of much lower quality than the RV measurements. 
If we found bisector variations with the RV period, then the planet hypothesis is refuted. 
On the other hand, a lack of bisector variations is not sufficient proof of the existence of the  planet. 
It could well be that a phenomenon produces measurable RV variations, but small bisector variations that are difficult to measure.

The stars also seem to show a lack of variations in the \textit{HIPPARCOS} photometry. 
Only one star, HD~106574, shows a weak peak in the L-S periodogram at the RV period. 
Again, this peak seems to be of low significance. 
However, the lack of photometric variations is only suggestive as the \textit{HIPPARCOS}  measurements were not contemporaneous to our data.

HD~44385, HD~106574, and HD~118904 have larger RV scatters than those predicted by the \citet{1995A&A...293...87K} relationship. 
Later spectral type stars, such as HD~106574 and HD~118904, show larger RV scatters than those given in \citet{2005PASJ...57...97S,2006A&A...454..943H}.
\citet{2012A&A...548A.118L} also have detected an exoplanet with similar orbital parameters and rms of the RV residuals.
Some more exoplanets discovered using BOES have shown large RV scatters \citep{2013A&A...549A...2L,2014A&A...566A..67L,2014JKAS...47...69L,2015A&A...584A..79L,2015A&A...580A..31H,2018AJ....155..120H}.

RV scatter in giant stars may have its origin not only from the stellar pulsations but also from the stellar activities.
\citet{1993ApJ...413..339H} found RV variations in $\alpha$ Boo with a period of 233 d and an amplitude of  $\sim$200 m s$^{-1}$. 
This RV period was the same period found in the He I 10830 variations in this star by \citet{1987ApJS...65..255L}. 
He I 10830 is a chromospheric activity indicator.
In this case, large RV variations are clearly due to the stellar activity.

Activity in giant stars is poorly understood and it may be that these are often not accompanied by variations in the ``classic'' indicators of stellar activity. 
We do not know the timescales or RV amplitudes of such activity jitter for these stars, which may have contributed to the RV scatters seen in some of our stars.
The exact cause of large RV scatters seen in our stars is yet to be understood by further study.

Rotating stars with surface features can also exhibit periodic RV variations which can be mimic  a ``planetary like'' signal. 
Our stars appear to be relatively inactive as shown by an absence of emission in the core of Ca II H lines.
Another check on rotational modulation can come from estimates of the rotational period of the star.

From $v_{\rm{rot}}$ sin $i$ and  $\textit{$R_{\star}$}$ (Table~\ref{tab:ste}) we estimate upper limits 
on the rotational period of 354.1 $\pm$ 72.0 days for \mbox{HD 44385}, 444.7 $\pm$ 122.0 days for \mbox{HD 97619}, 508.9 $\pm$ 152.1 days for \mbox{HD 106574}, 624.0 $\pm$ 265.7 days for \mbox{HD 118904}, 629.6 $\pm$ 125.7 days for \mbox{HD 164428}, and 267.4 $\pm$ 64.1 days for \mbox{HD 202432}.
\mbox{HD 118904} and  \mbox{HD 164428}  have estimated rotation periods comparable to the RV periods.
In the case of \mbox{HD 44385}, \mbox{HD 97619}, \mbox{HD 106574} and \mbox{HD 202432}
are the maximum rotational periods significantly larger than the RV periods. 
For these four stars our estimated rotational periods provide further support that we are not seeing rotational modulation.

Another explanation for the  RV variations in the six stars is a new, unknown form of stellar oscillations. 
One possibility is oscillatory convective modes. 
These have been proposed to explain the long-period variables \citep{2015MNRAS.452.3863S}, stars that are more evolved than the K giant stars of our work.
Coincidentally, only one of stars, HD 164428, is rather evolved with stellar radius larger than 20 $R_\odot$.
However, given that we know so little about long-period oscillations in K giant stars it seems that at the present time the most likely explanation for the RV variations in our stars is Keplerian motion by planetary companions. 

All of our targets are evolved intermediate mass stars. 
If the variations are due to planetary companions our detections add to the sample of exoplanets around stars more massive than the sun. 
Two of our stars have masses $\ge$ 1.5 $M_\odot$. 
Of the approximately 2200 exoplanets with good orbits and planet mass determinations, less than 5\% of the host stars have masses greater than  1.5 $M_\odot$ (source: exoplanets.org). 
Exoplanets around host stars with M $\ge$ 2 $M_\odot$ have a median $m$ sin $i$ of 2.7 $M_{J}$ and a median orbital period of $\sim$ 400 d.
Thus the companions to our six K giants have properties that are typical for planets around massive stars:
massive giant planets (1.9 -- 8.5 $M_{J}$) with orbital periods of several hundreds of days.

Although the SENS survey is not yet complete, at this stage we can still make  a rough estimate
of the planet frequency of our sample. 
We have found periodic variations in 31 of  our sample of 224 stars.
Among these, 17 stars have confirmed planets which is a planet occurrence rate of about 8\%. 
If all the RV variations are planetary  in nature, then the occurrence rate can be as high as $\approx$ 15\%. 
This is largely in line with the expectation that $\sim$10\% giant stars have planetary companions \citep{2015ApJ...798..112M}. 
A more detailed analysis of the statistics can be made once the SENS survey is completed.


\acknowledgments

This work is supported by the KASI (Korea Astronomy and Space Science Institute) through grant No. 2017-1-860-01. 
BCL acknowledges partial support by the KASI grant 2017-1-830-03.
Support for MGP and TYB was provided by the KASI under the R\&D program supervised by the Ministry of Science, ICT and Future Planning and by the National Research Foundation of Korea to the Center for Galaxy Evolution Research (No.2017R1A5A1070354). TYB was also supported by BK21 Plus of National Research Foundation of Korea.
S.G. would like to thank NSFC for the financial support through the grant No. U1531121.
This research made use of the SIMBAD database, operated at the CDS, Strasbourg, France. 
We thank BOAO for its generous support.

\bibliographystyle{apj}
\bibliography{reference}

\begin{table}[h!]
\renewcommand{\thetable}{\arabic{table}}
\centering
\caption{RV measurements for HD 44385. Table 3 is published in its entirety \\
in the electronic edition of the {\it Astronomical Journal}.  It \\
includes all six stars. Only HD 44385 is shown here for guidance \\
regarding its form and content.} \label{tab:rv1}
\begin{tabular}{ccc|ccc}
\tablewidth{0pt}
\toprule[1.5pt]
    JD       & $\Delta$RV   & $\pm$$\sigma$ &
    JD       & $\Delta$RV   & $\pm$$\sigma$ \\ 
-2450000  & m s$^{-1}$ & m s$^{-1}$ &
-2450000  & m s$^{-1}$ & m s$^{-1}$ \\
\midrule
5249.186772   &    -4.5   &    13.2 & 6739.992829  &     56.6  &     10.9 \\
5277.062617   &    -7.0   &    14.9 & 6922.262909  &    -95.2  &     11.2 \\
5554.202620   &    22.8   &    13.3 & 6965.959830  &    -56.6  &     10.9 \\
5843.317817   &   -10.6   &    11.1 & 7066.218427  &     77.5  &     13.1 \\
5934.119026   &   -40.3   &    10.2 & 7094.061052  &     75.4  &     16.3 \\
5963.063139   &   -80.1   &    11.0 & 7171.018942  &     49.5  &     18.4 \\
6024.980916   &   -87.6   &    11.6 & 7301.155470  &    -96.0  &     11.6 \\
6074.004223   &   -20.4   &    16.3 & 7330.319704  &   -124.3  &     12.7 \\
6210.342839   &    72.6   &    13.3 & 7378.167893  &   -112.0  &      9.7 \\
6258.199914   &    27.7   &    10.7 & 7491.026213  &     41.8  &     13.7 \\
6261.229862   &    31.9   &    10.3 & 7526.987711  &     78.5  &     11.9 \\
6287.184819   &    11.3   &    21.1 & 7529.996251  &     82.5  &     11.4 \\
6346.097590   &   -62.6   &    10.9 & 7704.047239  &     45.0  &     10.6 \\
6551.347025   &   111.9   &    13.9 & 7757.136001  &   -159.7  &     10.5 \\
6582.281632   &   104.7   &    15.4 & 7758.104706  &   -165.1  &     10.9 \\
6582.296436   &    96.6   &    10.8 & 7856.004581  &   -111.0  &      9.8 \\
6583.251127   &   132.0   &    12.6 & 8151.089358  &    -24.9  &     15.7 \\
6617.049357   &   137.9   &    10.8 &              &           &          \\
\bottomrule[1.5pt] 
\end{tabular}
\end{table}

\end{document}